\documentclass[preprint,aip,showpacs]{revtex4}

\usepackage{graphicx}

\usepackage{dcolumn} % Align table columns on decimal point

\usepackage{color}

\usepackage{bm}      % Bold math

\usepackage{tabularx}

\usepackage{float} %para que con [H] ponga las figs justo a continuacion

\begin{document}

\title{Energy spectrum, dissipation and spatial structures in reduced Hall magnetohydrodynamic}

\author{L.N. Martin$^1$, P. Dmitruk$^{1}$, D. O. Gomez$^{1,2}$}

\affiliation{$^1$ Departamento de F\'\i sica, Facultad de Ciencias Exactas y
             Naturales, Universidad de Buenos Aires and IFIBA, CONICET, Ciudad 
             Universitaria, 1428 Buenos Aires, Argentina. \\
             $^2$ Instituto de Astronomia y Fisica del Espacio, CONICET,
             Buenos Aires, Argentina}

\date{\today}

\begin{abstract}
We analyze the effect of the Hall term in the magnetohydrodynamic turbulence 
under a strong externally supported magnetic field, seeing how this changes 
the energy cascade, the characteristic scales of the flow  and the dynamics of 
global magnitudes, with particular interest in the dissipation.

Numerical simulations of freely evolving three-dimensional reduced magnetohydrodynamics
(RHMHD) are performed, for different values of the Hall parameter (the ratio of the ion
skin depth to the macroscopic scale of the turbulence) controlling the impact of the Hall
term. The Hall effect modifies the transfer of energy across scales, slowing down the
transfer of energy from the large scales up to the Hall scale (ion skin depth) and
carrying faster the energy from the Hall scale to smaller scales. The final outcome is an
effective shift of the dissipation scale to larger scales but also a development of smaller
scales. Current sheets (fundamental structures for energy dissipation) are affected in two
ways by increasing the Hall effect, with a widening but at the same time generating an
internal structure within them. In the case where the Hall term is sufficiently intense,
the current sheet is fully delocalized. The effect appears to reduce impulsive effects in
the flow, making it less intermittent.

\end{abstract}

\maketitle

\section{Introduction\label{intro}}

Among various kinetic corrections to magnetohydrodinamics models (MHD), the Hall effect
\cite{lib1,lib2} has been considered of particular importance in numerous studies: magnetic
reconnection \cite{Birn,Wang,Mo,Smi,Mor}, dynamo mechanisms \cite{Min}, accretion
disks\cite{Ac1,Ac2} and physics of turbulent regimes \cite{Dm,Min2,Gal,Dm2} are some of the
main examples. In this paper we want to study the general effect of the Hall term in
magnetohydrodynamic turbulence in plasmas embedded in a strong uniform magnetic field, through
numerical simulations. We studied the effect of this term on the dynamics of global magnitudes,
the cascade of energy, the characteristic scales and the intermittency of the flow.
 
The MHD models (one-fluid models) are important frameworks for the understanding of the large
scale dynamics of a plasma. However, these models fail to describe plasma phenomena with
characteristic length scales smaller than the ion skin depth $\rho_{ii}=c/\omega_{pi}$ (with
$\omega_{pi}$ the ion plasma frequency and $c$ speed of light). At this level, the Hall effect, 
which takes into account the separation between electrons and ions, becomes relevant. To describe
this regime is common to use the Hall MHD approximation, wich considers two-fluid effects through
a generalized Ohm's law which includes the Hall current. In presence of a strong external magnetic
field a new reduced model has been proposed, the RHMHD model\cite{Dm3,a,D3}). In this approximation, 
the fast compressional Alfv\'en mode is eliminated, while the shear Alfv\'en and the slow
magnetosonic modes are retained \cite{Za}. This new model (RHMHD) is an extension (including
the Hall effect) of the previously known reduced MHD (RMHD) model. The RMHD equations have
been used to investigate a variety of problems such as current sheet formation \cite{i12,i13},
non-stationary reconnection \cite{i14,i15}, the dynamics of coronal loops \cite{i16,i17}, and
the development of turbulence \cite{i18}. The self-consistency of the RMHD approximation has
been analyzed in ref. \cite{i19}. Moreover, numerical simulations have studied the validity
of the RMHD equations by directly comparing its predictions with the compressible MHD equations
in a turbulent regime \cite{i20}. In the same way it has been studied the validity of the RHMHD
model \cite{a}.

The control parameter (the Hall parameter) in this regime is $\epsilon=\rho_{ii}/L$, the
ratio of the ion skin depth $\rho_{ii}$ to the characteristic (large) scale of the turbulence
$L$. The influence of the Hall term can then be studied by increasing the Hall parameter in
different simulations.

The organization of the paper is as follows: Section \ref{equations} describes the sets of
equations used, and the codes to numerically integrate these equations. In Section \ref{resultados}
we present the numerical results: first a comparison between simulations with different Hall parameter
is performed for some global magnitudes, second we study the energy spectra, and then we use different
techniques to see the evolution of the characteristic structures of the flow. Finally, in Section
\ref{conclusiones} we list our conclusions.

\section{Equations and Numerical Simulations\label{equations}}

The compressible Hall MHD equations (dimensionless version) are 

\begin{eqnarray} \label{n-s}
\frac{\partial{\boldsymbol{V}}}{\partial t}= \boldsymbol{V} \times \boldsymbol{\omega} + \frac{1}{M_{A}^{2}}
\frac{\boldsymbol{J} \times \boldsymbol{B}}{\rho} - \nabla \left(\frac{\boldsymbol{V}^2}{2} +
\frac{\rho^{\gamma-1}}{M_{S}^2(\gamma-1)} \right)  \nonumber \\ + \nu \frac{\nabla^{2}{\boldsymbol{V}}}{\rho}
+ (\zeta+\frac{1}{3}\nu) \frac{\nabla ({\nabla \cdot \boldsymbol{V}})}{\rho},
\end{eqnarray}

\begin{equation} \label{continuidad}
\frac{\partial{\rho}}{\partial t}=-\nabla \cdot (\rho \boldsymbol{V}),
\end{equation}

\begin{equation} \label{M1}
\frac{\partial{\boldsymbol{A}}}{\partial t}={\boldsymbol{V}}\times \boldsymbol{B} -\epsilon \frac{\boldsymbol{J}
\times \boldsymbol{B}}{\rho} -\nabla \phi +\eta \nabla^2\boldsymbol{A},
\end{equation}

\begin{equation} \label{M2}
\nabla \cdot \boldsymbol{A}=0
\end{equation}

where $\boldsymbol{V}$ is the velocity field, $\boldsymbol{\omega}$ is the vorticity,
$\boldsymbol{J}$ is the current, $\boldsymbol{B}$ the magnetic field, $\rho$ is the density
of plasma, $\boldsymbol{A}$ and $\phi$ are the magnetic and electric potential. As indicated in Eq.
(\ref{n-s}) we used the barotropic law for the fluid, $p=cte.\rho^{\gamma}$ (here $p$ is
the pressure), in our case we consider $\gamma=5/3$. $M_S$ is the sonic Mach number, $M_A$ 
is the Alfven Mach number,  $\nu$ and $\zeta$ are the viscosities , $\eta$ is the resistivity
and $\epsilon=\rho_{ii}/L$ ($\rho_{ii}$ is the ion skin depth and $L$ the characteristic
scale of the turbulence) the Hall coefficient. All these numbers are control parameters in the 
numerical simulations. The Hall parameter $\epsilon$ appears in front of the Hall term in the
dimensionless equations, expressing the fact that the Hall term becomes important at scales 
smaller than the ion skin depth $\rho_{ii}$.

\subsection{RHMHD model}

The RHMHD model derived in the work by Gomez $et$ $al$ \cite{Dm3} and numerically tested by
Martin $et$ $al$ \cite{a} is a description of the two-fluid plasma dynamics in a strong
external magnetic field. The model assumes that the normalized (and dimensionless) magnetic field is of the form 
(the external field is along $\boldsymbol{\widehat{e}_z}$)

\begin{equation} \label{1}
\boldsymbol{B}=\boldsymbol{\widehat{e}_z}+\delta\boldsymbol{B}, \qquad  \mid\delta\boldsymbol{B}\mid\approx\alpha\ll 1,
\end{equation}

where $\alpha$ represents the typical tilt of magnetic field lines with respect to the
$\boldsymbol{\widehat{e}_z}$ direction, thus one expects

\begin{equation} \label{2}
\nabla_{\perp}\approx 1, \qquad  \partial_z\approx\alpha\ll 1.
\end{equation}

To ensure that the magnetic field $\boldsymbol{B}$ remains divergence free, it is assumed that

\begin{equation} \label{B}
\boldsymbol{B}= \boldsymbol{\widehat{e}_z} + \nabla \times(a \boldsymbol{\widehat{e}_z}
+ g \boldsymbol{\widehat{e}_x)}.
\end{equation}

The velocity field, in the more general case, can be decomposed as a superposition of a solenoidal
part (incompressible flow) plus the gradient of a scalar field (irrotational flow), i.e.

\begin{equation} \label{V}
\boldsymbol{V}=\nabla \times(\varphi \boldsymbol{\widehat{e}_z} + f \boldsymbol{\widehat{e}_x)}
+ \nabla \psi,
\end{equation}
where the potentials $a(\boldsymbol{r},t)$, $g(\boldsymbol{r},t)$,
$\varphi(\boldsymbol{r},t)$ and $f(\boldsymbol{r},t)$ are all assumed of
order $\alpha\ll1$ and $\psi(\boldsymbol{r},t)$ is of order $\alpha^2$ expressing the possibility a
slight compressibility (see details in \cite{a,Dm3,D3}). Introducing the expressions (\ref{B}) and (\ref{V}) in the compressible
set of equations (\ref{n-s})-(\ref{M2}) and taking the terms up to first and second order in
$\alpha$ the RHMHD model is obtained:

\begin{equation} \label{rhmhd1}
\frac{\partial \omega}{\partial t}= \frac{\partial j}{\partial z} + [j,a] - [\omega,\varphi]
+ \nu \nabla^2 \omega,
\end{equation}

\begin{equation} \label{rhmhd2}
\frac{\partial a}{\partial t}= \frac{\partial (\varphi-\epsilon b)}{\partial z} +
[\varphi,a] - \epsilon[b,a] + \eta \nabla^2 a,
\end{equation}

\begin{eqnarray} \label{rhmhd3}
\frac{\partial b}{\partial t}= \beta_p\frac{\partial (u-\epsilon j)}{\partial z} + [\varphi,b]
+ \beta_p[u,a]+\nonumber\\ - \epsilon \beta_p[j,a] + \beta_p\eta\nabla^2b,
\end{eqnarray}

\begin{equation} \label{rhmhd4}
\frac{\partial u}{\partial t}= \frac{\partial b}{\partial z} + [\varphi,u] - [a,b] +
\nu \nabla^2 u,
\end{equation}

where
     
\begin{eqnarray}
\omega=-\nabla^2_\perp \varphi\label{omega},\\
j=-\nabla^2_\perp a\label{j},\\
b=-\partial_y g\label{b},\\
u=-\partial_y f\label{u},
\end{eqnarray}

and the notation $[A,B]=\partial_x A \partial_y B - \partial_x B \partial_y A$ is employed.
$\beta_p=\beta\gamma/(1+\beta\gamma)$ is a function of the plasma ``$beta$''.

We use this set of equations to study how the Hall effect modifies the dynamics of magnetohydrodynamic
turbulence under a strong magnetic field.

\subsection{Numerical codes\label{codigos}}

We use a pseudospectral code to solve the set of equations (\ref{rhmhd1})-(\ref{rhmhd4}).
Periodic boundary conditions are assumed in all directions of a cube of side $2\pi L$ (where
$L\sim 1$ is the initial correlation length of the fluctuations, defined as the length unit).
In the codes, Fourier components of the fluctuations are evolved in time, starting from a
specified set of Fourier modes (see section \ref{resultados} for the specific initial conditions),
with given total energy and random phases. 

The same resolution is used in all simulations, $512^2$ in the perpendicular directions to
the external magnetic field and $32$ in the parallel direction (this is possible because the
structures that require high resolution only take place in the directions perpendicular to
the field), allowing four different runs to be done with four different Hall coeficients. 
The kinetic and magnetic Reynolds numbers are defined as $R=1/\nu$, $R_m=1/\eta$, based on
unit initial rms velocity fluctuation, unit length and non-dimensional values for the
viscosity and diffusivity. Here we used $R=R_m=1600$ ($\nu=\zeta=1/1600$, $\eta=1/1600$)
in all runs. We also considered a Mach number $M_S=1/4$, Alfven number $M_A=1$, and the Hall
coeficients $\epsilon$$=$$0$,  \,$1/32$,\,  $1/16$, and $1/8$ in all runs.

A second-order Runge-Kutta time integration is performed, the nonlinear terms are evaluated
using the standard pseudospectral procedure \cite{seudo}. The runs are freely evolved for $10$
time units (the initial eddy turnover time is defined in terms of the initial rms velocity
fluctuation and unit length). The magnetic field fluctuations
were less than ten percent of the external magnetic field value, so we are in the range of
validity of the RHMHD model.

\section{Results and discussion\label{resultados}}

We performed simulations for four different Hall coeficients, $\epsilon$$=$$0$,  \,$1/32$,\,  $1/16$, and $1/8$.

To generate the initial conditions, we consider initial Fourier modes (for magnetic and velocity
field fluctuations) in a shell in k-space $1 \le k \le 2$ at low wavenumbers, with constant
amplitude and random phases. Only plane-polarized fluctuations (transverse to the mean magnetic
field) are included, so these are (low- to high-frequency) Alfv\'en mode fluctuations and not
magnetosonic modes.  

The runs performed throughout this paper do not contain any magnetic or velocity stirring terms,
so the RHMHD system evolves freely.

\begin{figure}%[H]
\includegraphics[width=10.3cm]{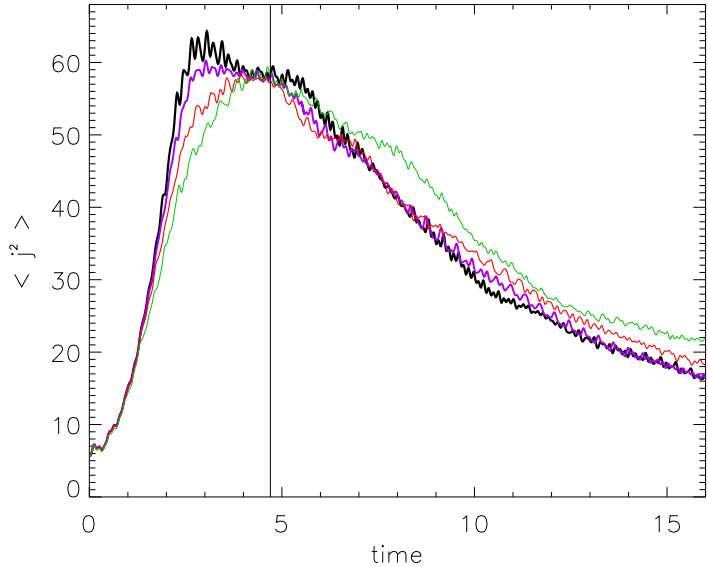}%[width=8.3cm]
\caption{\label{curr} Current density, $<J^2>$, as function of time for 
$\epsilon=0$, $1/32$, $1/16$ and $1/8$. The color of these curves are, black,
violet, red and green respectively (color online). The thickness of the lines decreases
proportionally to the value of $\epsilon$, thus, the thicker line corresponds
to $\epsilon=0$ and the finest line to $\epsilon=1/8$. The vertical straight
line indicates a particular time where all the scales have been developed in
all runs. Besides, in this time the value of $<J^2>$ is approximately the same
for all runs. This particular time will be used to study the different 
structures in the flows.}
\end{figure}

\begin{figure}%[H]
\includegraphics[width=10.3cm]{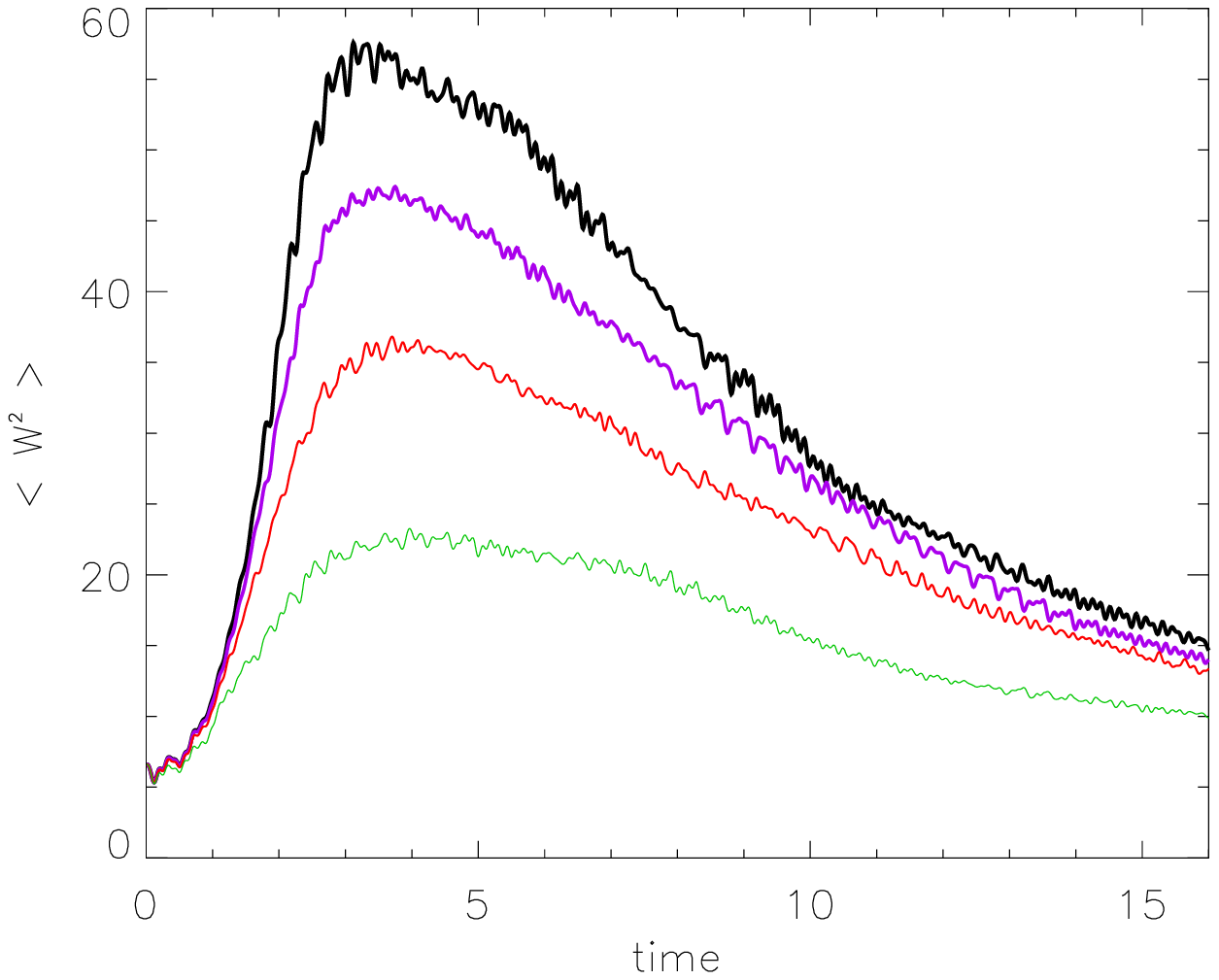}%[width=8.3cm]
\caption{\label{vor} Vorticity, $<\omega^2>$, as function of time for $\epsilon=0$,
$1/32$, $1/16$ and $1/8$. We are using the same convention of color/thickness than in figure 1}
\end{figure}

We study the influence of the Hall term in global quantities associated with the dissipation.
Figures \ref{curr} and \ref{vor} show the mean square current density $<J^2>$ and mean square
vorticity $<\omega^2>$ as function of time for  $\epsilon$$=$$0$,  \,$1/32$, \,$1/16$, \,$1/8$. Both $<J^2>$ and
$<\omega^2>$ show that as the Hall parameter is increased the dissipation decreases (in the case
of mean square vorticity this effect is considerably larger). Another remarkable effect is the
shift in the peaks of these functions: $<J^2>$ and $<\omega^2>$ take longer to reach its maximum  
with increasing $\epsilon$. The time of the peak indicates the time where all spatial scales
were developed (and therefore turbulence is fully developed). 

\begin{figure}%[H]
%\centerline{\includegraphics[width=8.3cm]{1}}
\includegraphics[width=10.3cm]{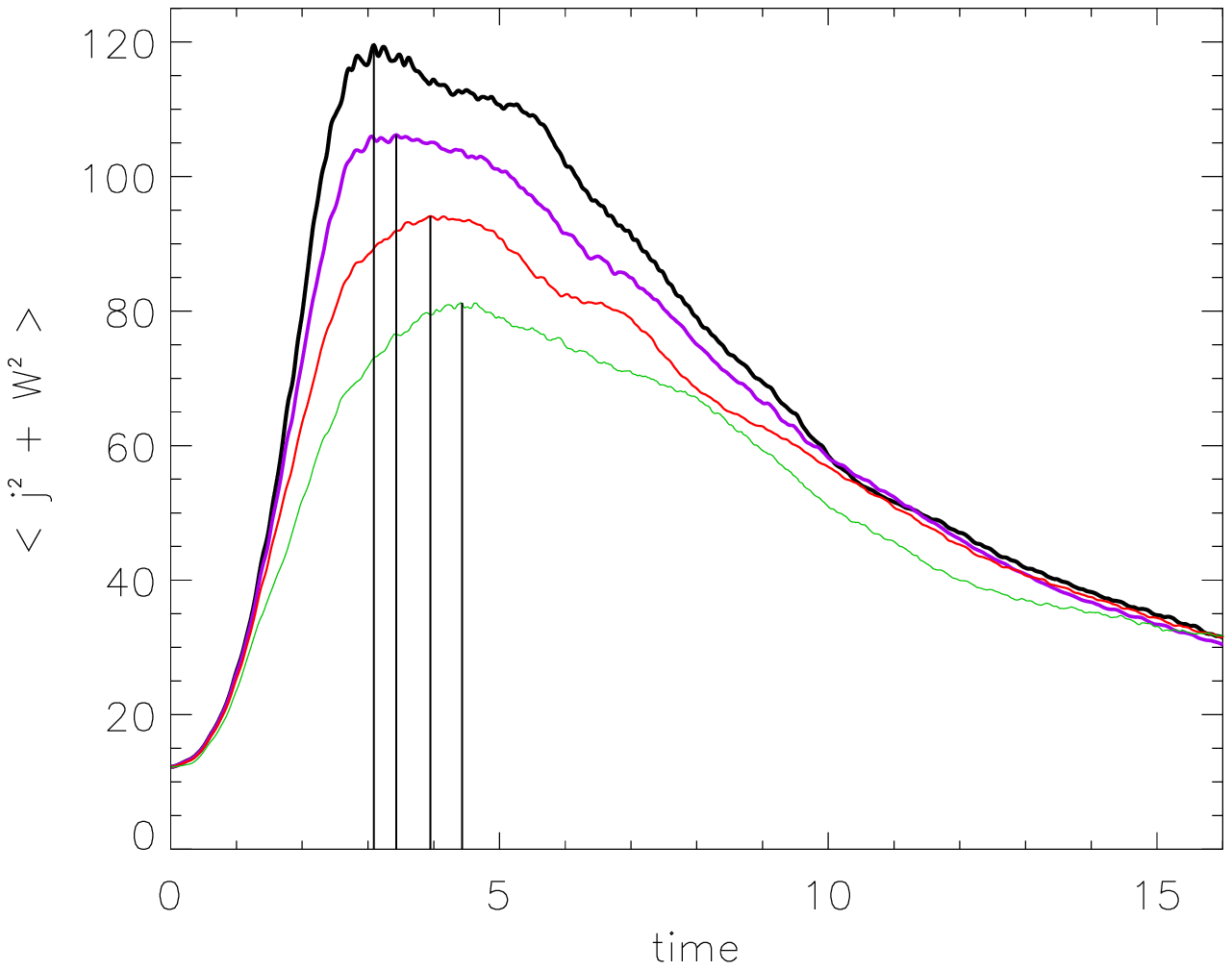}%[width=8.3cm]
\caption{\label{vorticity_current} $<J^2+\omega^2>$ as function of time for $\epsilon=0$,
$1/32$, $1/16$ and $1/8$. We are using the same convention of color/thickness than in figure 1.
The vertical straight lines indicate the maximum value for each curve.}
\end{figure}

Figure \ref{vorticity_current} shows  $<J^2+\omega^2>$ as function of time, the difference between
the peaks is more clear in this case. Here we see two effects that occur simultaneously as the Hall
coefficient is increased: The decrease in the dissipation and the delay in reaching the maximum point
(and hence the time that it takes to develop all the scales).  The first effect will have a direct
impact on the dissipation scale of the respective flows while the second shows how the Hall term
modifies its characteristic times.

It is relevant to note that the dissipation scale ($1/K_{diss}$) is related to the number of scales
that develop in the flow.  It is common to consider that the decrease in the dissipation scale
increases the range of developed scales in the flow (usually increases the size of the inertial
range). In the same manner, the increase in the scale of dissipation leads to a decrease in the
number of scales developed in the flow. However, it is not always this case. The results that we
will show below indicate that the Hall term affects the total width of the dissipation range
decreasing mildly the $K_{diss}$ (and therefore mildly increasing the dissipation scale) with
the increase of the $\epsilon$, at the same time the delays suffered by the dissipation peaks
is due to the development of a greater number of scales in the dissipative range due to a major
accumulation of energy in these scales. 

To quantify the dissipation scale (in Fourier space) of the different flows we use the conventional
criteria \cite{bisk} given by the equation (\ref{k_dis}).

\begin{equation} \label{k_dis}
K_{diss}= \left(\frac{<\boldsymbol{\omega}^2>+<\boldsymbol{J}^2>}{\nu^2}\right)^{1/4}
\end{equation}

In Table \ref{simulaciones} the Hall scale is shown along with the dissipation scale for each one
of the flows. Here we see the decrease of the $K_{diss}$ in quantitative form with the increase
of the Hall coefficient. Note that $K_{diss}<K_{max}=N/3=170$ means that the runs are marginally resolved
(see Wan et al 2010 \cite{wan} for more demanding requirements if higher order statistic is performed)

\begin{table}

\caption{\label{simulaciones} Hall and dissipation scales for the different runs.}

\begin{ruledtabular}

\begin{tabular}{cccc}

Run & $\epsilon$ &$K_{Hall}$ & $K_{diss}$\\ \hline
 1   &0        &            &132.25    \\ 
 2   &1/32     &32         &125.40    \\ 
 3   &1/16     &16         &124.58    \\
 4   &1/8      &8          &120.08    \\ 

\end{tabular}

\end{ruledtabular}

\end{table}

The decrease of the global dissipation with the Hall parameter and the increase in the time
of the peak development can not be understood by looking only at the temporal evolution of
the global magnitudes. These effects could be due to a change in the characteristic time of
the energy flow or to the development of small scale structures.

To better understand these issues we study the energy spectra and the size and shape of the
structures generated in the four runs. This help us to see whether or not the Hall effect
produces the development of small scales and also to understand this dynamic in terms of
the energy distribution.

\begin{figure}%[H]
\includegraphics[width=10.3cm]{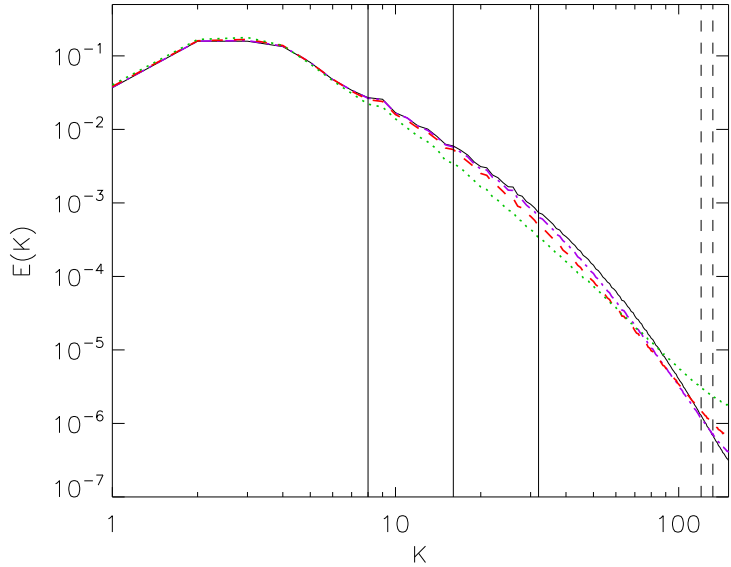}%[width=8.3cm]
\caption{\label{spectros} Energy spectra for $\epsilon=0$ (solid line), 
$\epsilon=1/32$ (dash-dotted), $\epsilon=1/16$ (dash) and $\epsilon=1/8$
(dotted). The vertical straight lines indicate the differents values of
$K_{Hall}=1/\epsilon$ for $\epsilon=1/32,1/16,1/8$. The vertical straight
dashed lines show the minimun and the maximum values of the $K_{diss}$
($\epsilon=1/8$ and $\epsilon=0$), through them we can see the effect
of the Hall term on the dissipation scale.}
\end{figure}

\begin{figure}%[H]
\includegraphics[width=10.3cm]{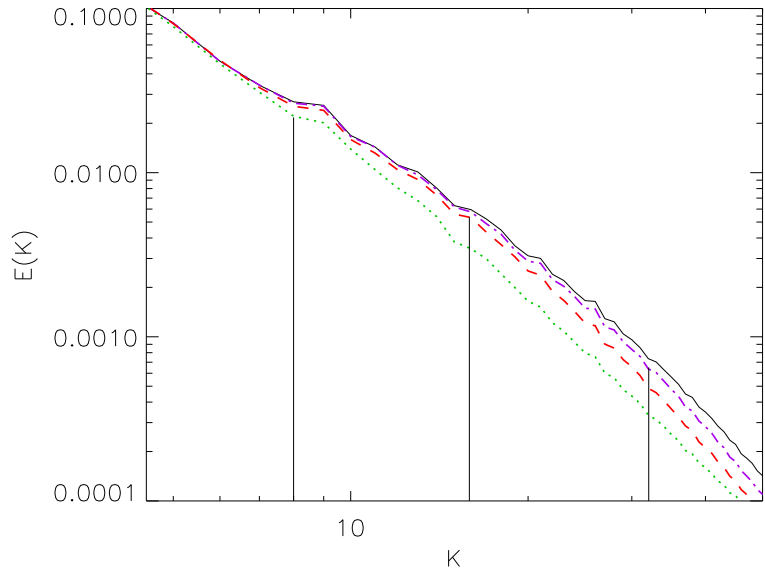}%[width=8.3cm]
\caption{\label{spectros_zoom} Enlarged view of a section of 
the energy spectra (limited $k$). The vertical straight lines
indicate the different values of $K_{Hall}=1/\epsilon$ for
$\epsilon=1/32,1/16,1/8$, these lines intersect each of the
corresponding curves.}
\end{figure}

Looking at the spectra we can see the distribution of the energy through different scales.
Figure \ref{spectros} compares the energy spectra for all runs and Figure \ref{spectros_zoom}
shows a zoom around of the Hall scales used.

As the Hall parameter is increased the energy spectrum is steeper at intermediate scales
preceeding the dissipation range. At the same time there is an increase in the energy on
scales smaller (larger $k$) than the dissipation scale (see Figs. \ref{spectros} and
\ref{spectros_zoom}). The effect of the Hall term is then twofold: first there is a slow down
of the energy transfer up to the Hall scale, resulting in a steeper spectrum, and then there
seems to be a driving of energy from the Hall scale up to the small scales (see \cite{Min3}
for a study of how the Hall term affects the transfer of energy at different scales). 
A shift of the effective dissipation scale to larger scales is then to be expected (as indicated
by the values of $K_{diss}$ given before) as well as a decrease in the global dissipation values.
At the same time, since the Hall term increases the number of effective scales on which the
dynamics occurs (as evindenced by the extended spectra at small scales) a longer time to reach
the peak of dissipation is expected, as previously shown.
 
\begin{figure}%[H]
\includegraphics[width=8.5cm]{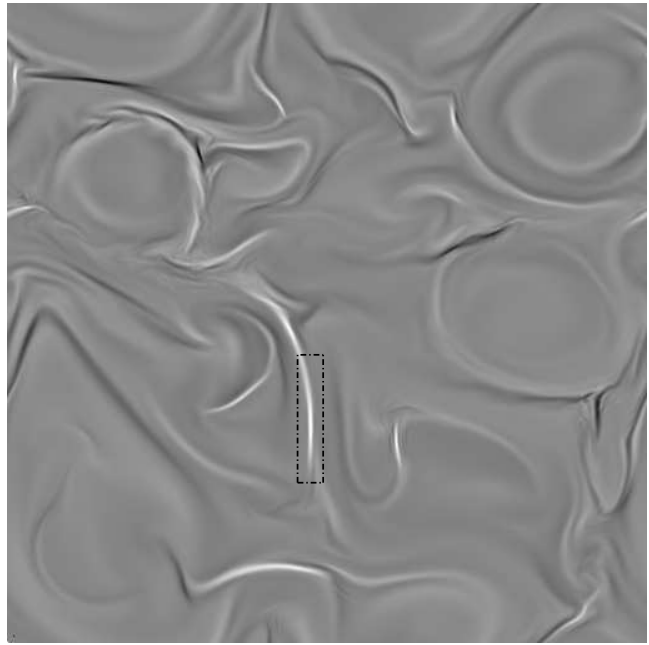}%[width=7cm]
\caption{\label{corte_1} Parallel component of the current density in a 
perpendicular plane to the external magnetic field in the case with
$\epsilon=0$ .Tones indicate out of plane current, with light tones=positive
and dark tones=negative.}
\end{figure}

\begin{figure}%[H]
\includegraphics[width=8.5cm]{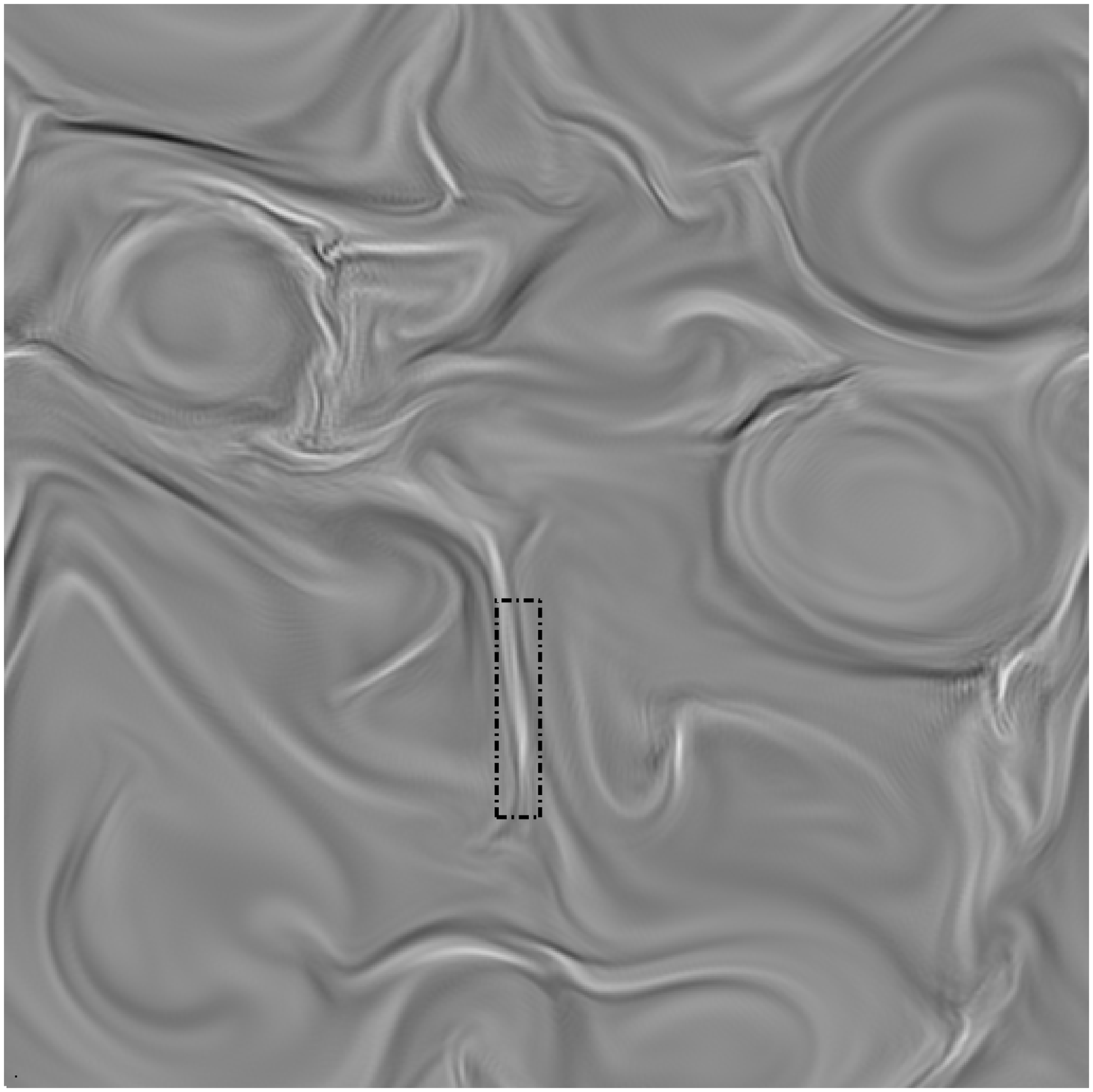}%[width=7cm]
\caption{\label{corte_2} Parallel component of the current density 
in a perpendicular plane to the external magnetic field in the case 
with $\epsilon=1/32$ .Tones indicate out of plane current, 
with light tones=positive and dark tones=negative.}
\end{figure}

\begin{figure}%[H]
\includegraphics[width=8.5cm]{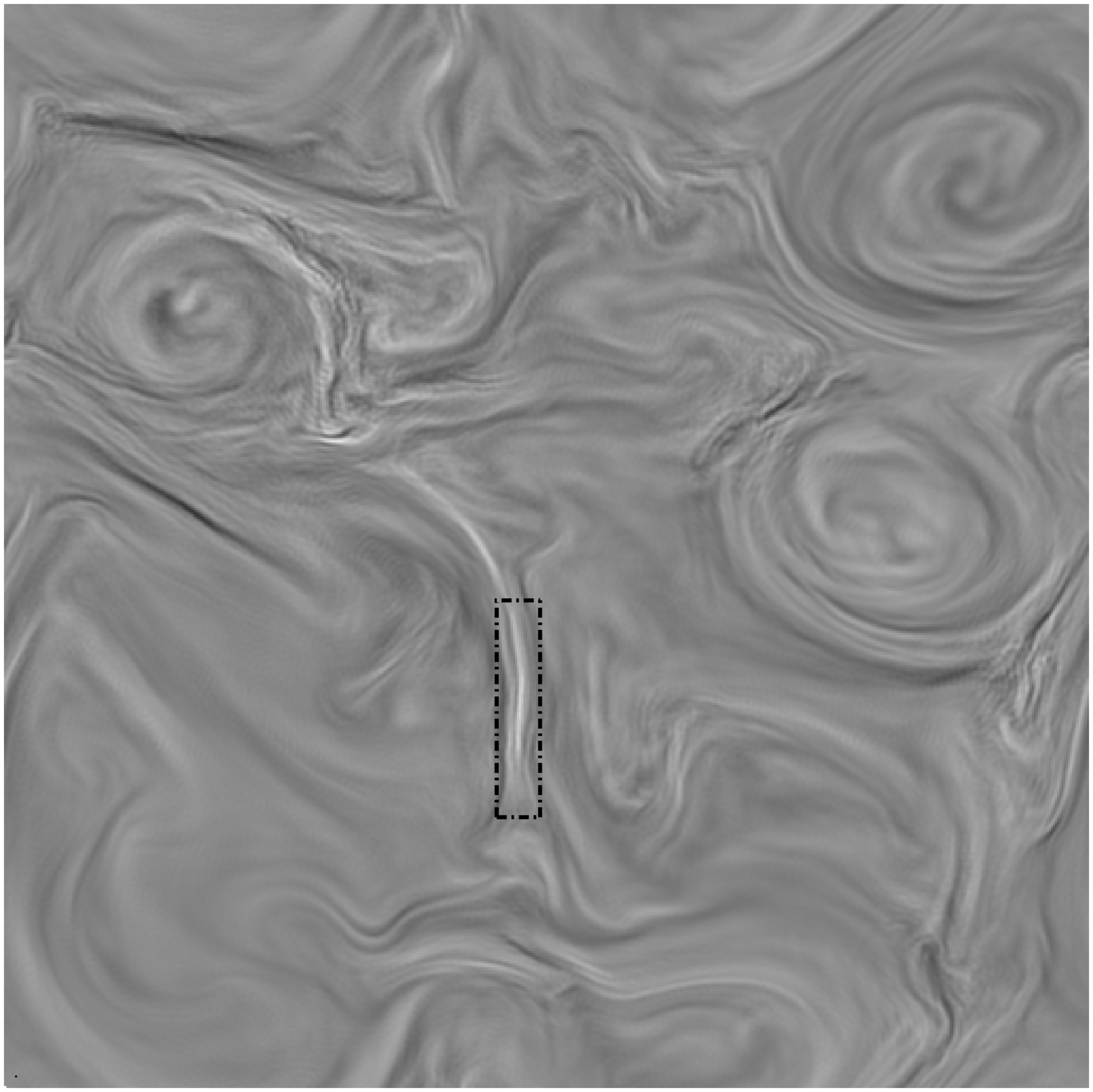}%[width=7cm]
\caption{\label{corte_3} Parallel component of the current density in 
a perpendicular plane to the external magnetic field in the case 
with $\epsilon=1/16$ .Tones indicate out of plane current, 
with light tones=positive and dark tones=negative.}
\end{figure}

\begin{figure}%[H]
\includegraphics[width=8.5cm]{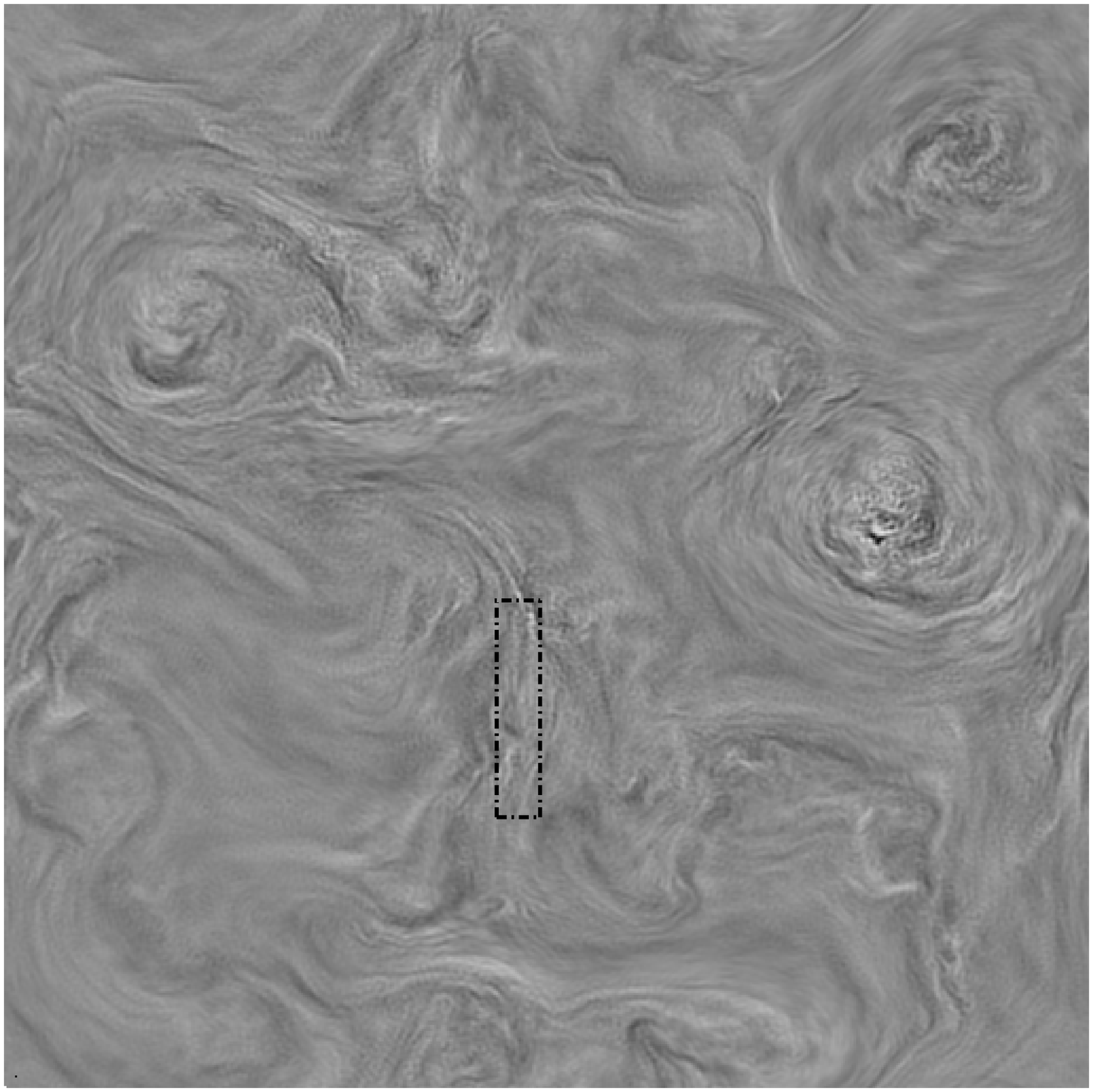}%[width=7cm]
\caption{\label{corte_4} Parallel component of the current density in a 
perpendicular plane to the external magnetic field in the case 
with $\epsilon=1/8$ .Tones indicate out of plane current, 
with light tones=positive and dark tones=negative.}
\end{figure}

We study the characteristic structures of the flow and the effect of the Hall term by looking at
the current density field. Figures \ref{corte_1}-\ref{corte_4} show the parallel component of
the current density in a perpendicular plane to the external magnetic field at a given time for
the different runs. The time was chosen in which all scales have been developed for all the
flows (this time is indicated in Figure \ref{curr}). Also, for this particular time, the value
of $<J^2>$ is approximately the same for all the runs.

In Figure \ref{corte_1} ($\epsilon=0$) we can clearly distinguish the current sheets
that form in the flow. We have highlighted one of the current sheets with a rectangle
with dashed lines. This structure is localized and well defined. Looking at the change
of this structure with the value of Hall parameter we can see two effects: first,
a widening of the sheet and secondly an internal filamentation. The widening is very
clear from Figure \ref{corte_1} with $\epsilon=0$ to Figure \ref{corte_2} with
$\epsilon=1/32$ and the internal filamentation starts to be seen in the Figure
\ref{corte_3}, with $\epsilon=1/16$, where also the thickness has increased. In the
case with higher $\epsilon=1/8$ the current sheet is completely filamentated, and is
hard to distinguish a clear structure at all.

These results are complementary to the results observed in the spectra and global
magnitudes and corroborate the idea that the Hall effect results in an effective shift
of the  dissipation scale (current sheet thickness getting larger) but also an increase
in the dynamical scale range (increase of filamentation). 

\begin{figure}%[H]
\includegraphics[width=10.3cm]{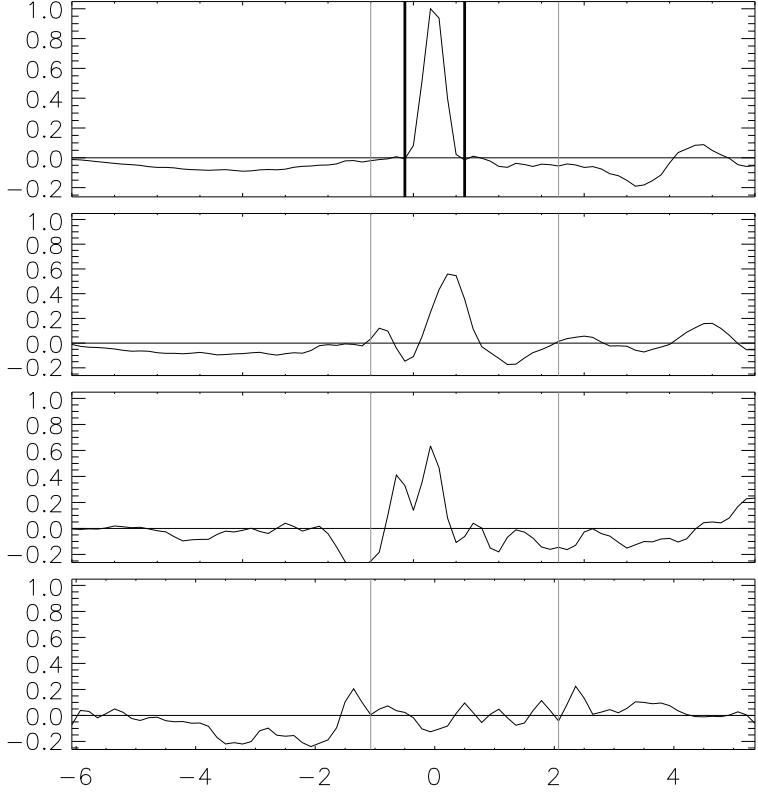}%[width=8.3cm]
\caption{\label{perfiles} Current density profile of the current sheets 
studied. From top to bottom are the cases with $\epsilon=0$, $\epsilon=1/32$, 
$\epsilon=1/16$, and $\epsilon=1/8$. The net flow of current (the absolute value) is
the same within the clear lines (vertical outside lines), and the dark lines
mark the original position of the current sheet when $\epsilon=0$.} 
\end{figure}

To better quantify the effect we have just observed, we plot the profile of the current
density in the direction perpendicular to the current sheet seen in the figures
\ref{corte_1}-\ref{corte_4}. These profiles are shown in Figure \ref{perfiles}. 

The net flow of current (the absolute value) is the same within the clear lines (vertical outside lines).
When $\epsilon=0$ the current sheet is perfectly located (the dark lines mark the original
position of the current sheet when $\epsilon=0$) and it is homogeneous (in the sense that
we have a single well defined peak). When $\epsilon=1/32$ the original sheet expands and two
sheets or filaments appear in their place (there are now two peaks). For $\epsilon=1/16$ the
width of the main sheet is greater and there is now a clear internal structure. In this case
the ambiguity that arises is whether we have one or more sheets of current (compare Figure
\ref{perfiles} with \ref{corte_3}) and hence the ambiguity of whether we have a wider sheet
or two thin sheets. When $\epsilon=1/8$ there is no trace of the current sheet. 

At this point we should make an important observation about the evolution of current sheets
as a function of the Hall parameter. As we saw there are two effects acting simultaneously,
the widening of what could be considered the overall structure of the sheet and the internal
filamentation that this suffers. In this way it could be interpreted that the Hall effect widens 
the current sheets (if we see the entire structure like the sheet) or on the other hand the
Hall effect produces finer sheets (considering that the small filaments are the sheets).
To remove the ambiguity (in semantics), we propose to speak in terms of dissipation, so if the
global structure dissipates less energy as we increase the Hall parameter we will say that 
the sheet is being widened, otherwise, if more energy is dissipated we will say that the
relation between size and intensity of internal filaments allow us to identify new current sheets.
Our results agree with the first frame of mind: as a function of dissipation the current sheets
are widening and even more when $\epsilon=1/8$ there is no trace of any structure that could
be identified as a current sheet.

\section{Conclusions\label{conclusiones}}

We performed numerical simulations of magnetohydrodynamic turbulence in strong magnetic fields,
including the Hall effect, and varying the Hall parameter.

We found that the Hall term affects the scales that are situated between the Hall scale and
the dissipation scale, resulting in a decrease in the accumulation of energy in this scale
range. The result is an effective shift of the dissipation scale but also a transfer of
energy to smaller scales. When the separation between the Hall scale and the dissipation
scale is larger an increasingly sharp steepening of the energy spectrum occurs at this
range of scales. The final outcome is the generation of smaller scales when the Hall scale
increases.

Localized structures are destroyed by this effect, suffering a gradual filamentation with
the increase of the Hall scale. The latter effect is manifested, for example, in the widening 
of the current sheets and the formation of internal structures within the sheets. At the same
time a decrease of the total energy dissipated is observed.

The results presented here suggest that the Hall effect reduces the intermittency, however
a more detailed study of this property should be performed. We defer this to further work. 

\acknowledgments
Research supported by grants UBACYT 20020090200602, 
PICT 2007-00856 and 2007-02211
from ANPCyT and PIP 11220090100825 from CONICET. 
We acknowledge the Marie Curie Project FP7 PIRSES-2010-269297 - ``Turboplasmas''.
The authors would like to acknowledge comments by P. D. Mininni that helped 
them substantially improve this work.

\end{document}